# Automated Text Summarization Base on Lexicales Chain and graph Using of WordNet and Wikipedia Knowledge Base

**Mohsen Pourvali and Mohammad Saniee Abadeh**

**Department of Electrical & Computer Qazvin Branch Islamic Azad University**
Qazvin, Iran

**Department of Electrical and Computer Engineering at Tarbiat Modares University**
Tehran, Iran

**Abstract**

The technology of automatic document summarization is maturing and may provide a solution to the information overload problem. Nowadays, document summarization plays an important role in information retrieval. With a large volume of documents, presenting the user with a summary of each document greatly facilitates the task of finding the desired documents. Document summarization is a process of automatically creating a compressed version of a given document that provides useful information to users, and multi-document summarization is to produce a summary delivering the majority of information content from a set of documents about an explicit or implicit main topic. The lexical cohesion structure of the text can be exploited to determine the importance of a sentence/phrase. Lexical chains are useful tools to analyze the lexical cohesion structure in a text .In this paper we consider the effect of the use of lexical cohesion features in Summarization, And presenting a algorithm base on the knowledge base. Our$^s$ algorithm at first find the correct sense of any word, Then constructs the lexical chains, remove Lexical chains that less score than other ,detects topics roughly from lexical chains, segments the text with respect to the topics and selects the most important sentences. The experimental results on an open benchmark datasets from DUC01 and DUC02 show that our proposed approach can improve the performance compared to sate-of-the-art summarization approaches.

## 1. Introduction

The technology of automatic document summarization is maturing and may provide a solution to the information overload problem. Nowadays, document summarization plays an important role in information retrieval (IR). With a large volume of documents, presenting the user with a summary of each document greatly facilitates the task of finding the desired documents. Text summarization is the process of automatically creating a compressed version of a given text that provides useful information to users, and multi-document summarization is to produce a summary delivering the majority of information content from a set of documents about an explicit or implicit main topic [14]. Authors of the paper [10] provide the following definition for a summary: "A summary can be loosely defined as a text that is produced from one or more texts that conveys important information in the original text(s), and that is no longer than half of the original text(s) and usually significantly less than that. Text here is used rather loosely and can refer to speech, multimedia documents, hypertext, etc. The main goal of a summary is to present the main ideas in a document in less space. If all sentences in a text document were of equal importance, producing a summary would not be very effective, as any reduction in the size of a document would carry a proportional decrease in its in formativeness. Luckily, information content in a document appears in bursts, and one can therefore distinguish between more and less informative segments. Identifying the informative segments at the expense of the rest is the main challenge in summarization". assumes a tripartite processing model distinguishing three stages: source text interpretation to obtain a source representation, source representation transformation to summary representation, and summary text generation from the summary representation. A variety of document summarization methods have been developed recently. The paper [4] reviews research on automatic summarizing over the last decade. This paper reviews salient notions and developments, and seeks to assess the state-of-the-art for this challenging natural language processing (NLP) task. The review shows that some useful summarizing for various purposes can already be done but also, not surprisingly, that there is a huge amount more to do. Sentence based extractive summarization techniques are commonly used in automatic summarization to produce extractive summaries. Systems for extractive summarization are typically based on technique for sentence extraction, and attempt to identify the set of sentences that are most important for the overall understanding of a given document. In paper [11] proposed paragraph extraction from a document based on intra-document links between paragraphs. It yields a text relationship map (TRM) from intra-links, which indicate that the linked texts are semantically related. It proposes four strategies from the TRM: bushy path, depth-first path, segmented bushy path, augmented segmented bushy path.

In our study we focus on sentence based extractive summarization. In this way we to express that The lexical cohesion structure of the text can be exploited to determine the importance of a sentence. Eliminate the ambiguity of the word has a significant impact on the inference sentence. In this article we will show that the separation text into the inside issues by using the correct concept Noticeable effect on the summary text is created. The experimental results on an open benchmark datasets from DUC01 and DUC02 show that our proposed approach can improve the performance compared to state-of-the-art summarization approaches.

The rest of this paper is organized as follows: Section 2 introduces related works, Word sense disambiguation is presented in Section 3, clustering of the lexical chains is presented in Section 4, text segmentation base on the inner topics is presented in Section 5, The experiments and results are given in Section 6. Finally conclusion presents in section 7.

## 2. Related work

Generally speaking, the methods can be either extractive summarization or abstractive summarization. Extractive summarization involves assigning salience scores to some units (e.g.sentences, paragraphs) of the document and extracting the sentences with highest scores, while abstraction summarization (e.g.http://www1.cs.columbia.edu/nlp/newsblaster/) usually needs information fusion, sentence compression and reformulation [14].

Sentence extraction summarization systems take as input a collection of sentences (one or more documents) and select some subset for output into a summary. This is best treated as a sentence ranking problem, which allows for varying thresholds to meet varying summary length requirements. Most commonly, such ranking approaches use some kind of similarity or centrality metric to rank sentences for inclusion in the summary – see, for example, [1].The centroid-based method [3] is one of the most popular extractive summarization methods. MEAD (http://www.summarization.com/mead/) is an implementation of the centroid-based method for either single-or-multi-document summarizing. It is based on sentence extraction. For each sentence in a cluster of related documents, MEAD computes three features and uses a linear combination of the three to determine what sentences are most salient. The three features used are centroid score, position, and overlap with first sentence (which may happen to be the title of a document). For single-documents or (given) clusters it computes centroid topic characterizations using tf–idf-type data. It ranks candidate summary sentences by combining sentence scores against centroid, text position value, and tf–idf title/lead overlap. Sentence selection is constrained by a summary length threshold, and redundant new sentences avoided by checking cosine similarity against prior ones. In the past, extractive summarizers have been mostly based on scoring sentences in the source document. In paper [12] each document is considered as a sequence of sentences and the objective of extractive summarization is to label the sentences in the sequence with 1 and 0, where a label of 1 indicates that a sentence is a summary sentence while 0 denotes a non-summary sentence. To accomplish this task, is applied conditional random field, which is a state-of-the-art sequence labeling method .In paper [15] proposed a novel extractive approach based on manifold–ranking of sentences to query-based multi-document summarization. The proposed approach first employs the manifold–ranking process to compute the manifold–ranking score for each sentence that denotes the biased information-richness of the sentence, and then uses greedy algorithm to penalize the sentences with highest overall scores, which are deemed both informative and novel, and highly biased to the given query. The summarization techniques can be classified into two groups: supervised techniques that rely on pre-existing document-summary pairs, and unsupervised techniques, based on properties and heuristics derived from the text. Supervised extractive summarization techniques treat the summarization task as a two-class classification problem at the sentence level, where the summary sentences are positive samples while the non-summary sentences are negative samples. After representing each sentence by a vector of features, the classification function can be trained in two different manners [7]. One is in a discriminative way with well-known algorithms such as support vector machine (SVM) [16]. Many unsupervised methods have been developed for document summarization by exploiting different features and relationships of the sentences – see, for example [3] and the references therein. On the other hand, summarization task can also be categorized as either generic or query-based. A query-based summary presents the information that is most relevant to the given queries [2] and [14] while a generic summary gives an overall sense of the document's content [2] , [4] , [12] , [14]. The QCS system (Query, Cluster, and Summarize) [2] performs the following tasks in response to a query: retrieves relevant documents; separates the retrieved documents into clusters by topic, and creates a summary for each cluster. QCS is a tool for document retrieval that presents results in a format so that a user can quickly identify a set of documents of interest. In paper [17] are developed a generic, a query-based, and a hybrid summarizer, each with differing amounts of document context. The generic summarizer used a blend of discourse information and information obtained through traditional surface-level analysis. The query-based summarizer used only query-term information, and the hybrid summarizer used some discourse information along with query-term information. The article [18] presents a multi-document, multi-lingual, theme-based summarization system based on modeling text cohesion (story flow).

## 3. Word Sense Disambiguation

For extracting lexical chains in a document, all words and correct senses of these words should be known. Humans disambiguate words by the current context. Lexical chaining algorithms depend on an assumption, and this assumption is that correct sense of words has stronger relations with other word senses. Using this assumption, lexical chaining algorithms first try to disambiguate all word occurrences. For this reason, word sense disambiguation (WSD) is an immediate application of lexical chains and an extrinsic evaluation methodology.

3.1 generating and traversing the WordNet graph

The algorithm presented in this paper is based on lexical chains therefore the system needs to deeply analyze the text. Per word has a sense based on it's position in the sentence. For instance, the word **bank** in the follow sentences has different senses:"Beautiful bank of river" and "Bank failures were a major disaster".   In first sentence bank means river's coast, but in the second sentence it means economic bank. The most appropriate sense must be chosen for this word and it cause increasing the connectedness in a lexical chain. In the algorithm presented in this paper , word sense are calculated locally . in this way the best word sense is extracted .we also use WordNet as an external source for disambiguation let $w_i$ be a word in the document ,and $w_i$ have n senses $\{w_{i_1}, w_{i_2}, \ldots, w_{i_k}, \ldots, w_{i_n}\}$.in this procedure for finding the meaning of two words related locally together and placed in the same sentence , we assume all of the possible meanings and senses of per word as the first level of the traversing word tree then we process every sense in a returning algorithm .Next , we connect all the relations for that sense as it's  descendants ,and these descendants are generated through relations that are Hypernym ,... . We do this process in a returning manner for n levels. Next, every first level sense of the one word compare with all the first level senses of the other word .Afterwards, the numbers of equalities are considered in integer digit .the same comparison is done for another word .if there isn't any equality, for each word we choose first sense that is most common.

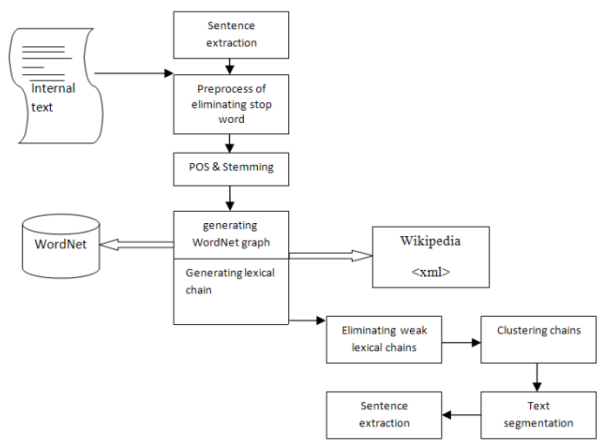

Fig. 1  Diagram of algorithm's steps

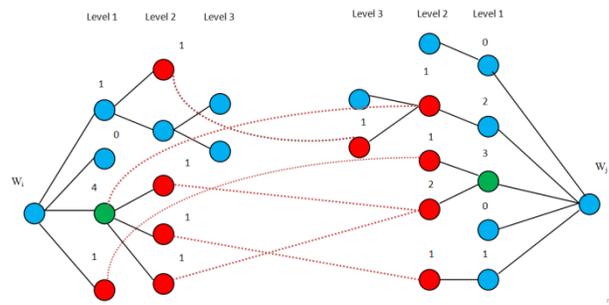

Fig. 2 Sample graph built on the 2 words

 In the above figure, we illustrate the relations of the tree .the root of the tree is considered as the target word, and the first level nodes as the senses of the target words. The nodes of the second, third,...levels are senses related with the first level nodes with Hypernym ,… relations. This tree is generated using returning functions and traversing of the tree is in the returning manner.

```
Function Hyp(ref Node t,int level)
       string[] sp
       for  i = 0  to  EndOfFile(wn_hyp) do
         ReadLine_From_File(wn_hyp)
         sp=Split_String_base_of('(', ',', ')')
       if t.index == sp[1]
            tnew=Create New Nod(sp[2])
            Call Hyp(ref tnew,level-1)
            Add_New_Nod_ToList(tnew)
       end if
end for
```

Fig. 3 Algorithm for creation WordNet graph

The above algorithm is one of the functions used for producing WordNet graph .this function is the part of the graph related with Hypernym relation .We use the great encyclopedia of Wikipedia because of the lack of special names in knowledge base of WordNet. This is done using the 3.5G XML file that is downloaded from dumps.wikipedia.org site. We have created a Xml_Reader for this file, and then goal word abstract is extracted. Extracted abstract is used same of the Glosses of another sentence's word we use creating the graph and traversing of it just for the first ,middle ,and last sentences ,and it is useful because these sentences usually encompass concise expression of the concept of the paragraph in most of the documents .in this manner we decrease the space of interpretation and therefore the time of calculation and the space of memory because we just need to keep some highlight sentences related with each other. After clarifying the actual senses of the all words in the prominent sentences and with the similarities and relations between every pair of the words, we put them in incessant lexical chains. For example in the tree of two words, and through the traversing of the first word, we put these two words in the same lexical chain as soon as we reach the first common sense between the subordinate graph of the first word and the first level nodes of the second word .For each lexical chain $LC_i$ , $w_3^1$ symbolizes that this word occur in the first sentence and the third sense of this word is chosen as the best sense. lexical chains created at first are generated from highlight sentences, and we use different algorithm for putting other words of sentences in the relevant lexical chains. in this algorithm with some changes in Lesk algorithm ,we use gloss concepts to represent similarities and differences of two words. let w1 , w2 are two words in text .firstly we extract senses of per word in normal Lesk algorithm from knowledge base

$$s1 \in sense\ (w1)\ and\ s2 \in sense\ (w2) \qquad (1)$$

then we find overlaps between gloss concepts

$$score_{lesk}(s1,s2) = |gloss(s1) \cap gloss(s2)| \qquad (2)$$

And every two concepts that have more similarities are chosen as the target words. Moreover, we use not only uni-gram (sequence of one word) overlaps , but also bi-gram (sequence of two words) overlaps .if there is one of the senses the first word in gloss concepts of the second word, we give one special score to this two senses. We do this because two concepts may have common words that are not related with their similarities and it causes increasing in scores of that two senses and makes a mistake in choosing related word as a result. Considering the word sense in gloss concept of the second word's sense, we can award an additional chance to this sense to be chosen in process of choosing words for chains from words that are not semantically related in fact.

$$if\bigl(s1 \in gloss\ (s2)\ or\ s2 \in gloss\ (s1)\bigr) \qquad (3)$$

$$score\ (s1,s2) = score_{lest}(s1,s2) + \lambda$$

λ is an additional score, and considering average existed words in sense's gloss concept and experimental tests, we find that the best value for λ is 5 . it is important in surveying gloss concepts to survey just existed names and existed verbs. At first, there are lexical chains generated from highlight sentences with traversing the graph, and with assuming $LC_i$ as one of the lexical chains generated from last step and $W_j$ as one of the other sentence's words and with using the above algorithm , $W_j$ is compared with members of lexical chain $LC_i$ .if the similarity's score of $W_j$ with one of the members of $LC_i$ is more than threshold T , $W_j$ is added to $LC_i$ and from now on, other residual words are investigated  based on their similarities with members of $LC_i$ and $W_j$ ,too.

```
Function (Word₁,Word₂)
H=0 , WordInGloss = 0
For i=0 to CountOfSenseWord₁
 For j=0 to CountOfSenseWord₂
  For s=0 to CSG₁[i]
   For k=0 to 1
    If s+k == s
     N = WSG₁[s]
    elseIf s <> CSG₁[i]
     n = WSG₁[s] + " " + WSG₂[s + k]
    else break
    if GlossWord₂[j].Contains(n)
     H++
    End if
   End for
  End for
  If GlossWord₂[j].Contains(Word₁) or
  GlossWord₂[i].Contains(Word₁)
   WordInGloss = 5
  End if
  F = H + WordInGloss
  ed = new edge(SenseWord₁[i], SenseWord₂[j], f)
  AllEdge.Add(ed)
 End for
End for
```

Fig. 4 Compare algorithm for Glosses

## 4. Clustring Lexical Chains

After lexical chains are constructed for the text, there will be some weak lexical chains formed of single word senses.
For each lexical chain $LC_i$, a sentence occurrence vector $V_i$ is formed. $v_i = \{s_{1_i}, \dots, s_{k_i}, \dots, s_{n_i}\}$ where n is the number of sentences in the document. Each $s_{k_i}$ is the number of $LC_i$ members in the sentence $k$. If sentence $k$ has 3 members of $LC_i$ then $s_{k_i}$ is 3. Two lexical chains $LC_i$ and $LC_j$ go into the same cluster if their sentence occurrence vectors $V_i$ and $V_j$ are similar.

Our clustering algorithm, starts from an initial cluster distribution, where each lexical chain is in its own cluster. Thus, our clustering algorithm starts with n clusters, where n is the number of lexical chains. Iteratively the most similar cluster pair is found and they are merged to form a single cluster. Clustering stops when the similarity between the most similar clusters is lower than a threshold value. for this purpose we used the well known formula from Linear Algebra:

$$Cos(\theta) = \frac{v_i + v_j}{||v_i|| \, ||v_j||} \qquad (4)$$

In the equation $||v_i||$ represents the Euclidean Length for the vector.

## 5. Sequence Extraction

In our algorithm, the text is segmented from the perspective of each lexical chain cluster, finding the hot spots for each topic. For each cluster, connected sequences of sentences are extracted as segments. Sentences that are cohesively connected are usually talking about the same topic. For each lexical chain cluster $Cl_j$, we form sequences separately. For each sentence $S_k$, if sentence $S_k$ has a lexical chain member in $Cl_j$, a new sequence is started or the sentence is added to the sequence. If there is no cluster member in $S_k$, then the sequence is ended. By using this procedure, text is segmented with respect to a cluster, identifying topic concentration points. Figure 5 is an example of Text Segmentation.

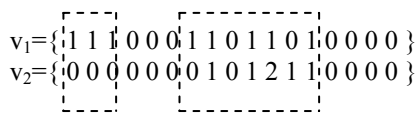

$v_1=\{1\;1\;1\;0\;0\;0\;1\;1\;0\;1\;1\;0\;0\;0\;0\;0\}$
$v_2=\{0\;0\;0\;0\;0\;0\;0\;1\;0\;1\;2\;1\;1\;0\;0\;0\;0\}$

Fig. 5 example of Text Segmentation

Each sequence is scored using the formula in Equation (5).

$$Score(Sequence_i) = Score(Cl_i) * l_i * \frac{(1 + SLC_i) * PLC_i}{f^2} \qquad (5)$$

Where $l_i$ is the number of sentences in the sequence$_i$. $SLC_i$ is the number of lexical chains that starts in sequence$_i$. $PLC_i$ is the number of lexical chains having a member in sequence$_i$ and $f$ is the number of lexical chains in cluster. Score of the cluster score($Cl_i$), is the average score of the lexical chains in the cluster. Our scoring function tries to model the connectedness of the segment using this cluster score.

## 6. Experiments and Results

In this section, we conduct experiments to test our summarization method empirically.

### 6.1 Datasets

For evaluation the performance of our methods we used two document datasets DUC01 and DUC02 and corresponding 100-word summaries generated for each of documents. The DUC01 and DUC02 are an open benchmark datasets which contain 147 and 567 documents-summary pairs from Document Understanding Conference (http://duc.nist.gov). We use them because they are for generic single-document extraction that we are interested in and they are well preprocessed. These datasets DUC01 and DUC02 are clustered into 30 and 59 topics, respectively. In those document datasets, stop words were removed using the stop list provided in ftp://ftp.cs.cornell.e-du/pub/smart/english.stop and the terms were stemmed using Porter's scheme [9], which is a commonly used algorithm for word stemming in English.

### 6.2 Evaluation metrics

There are many measures that can calculate the topical similarities between two summaries. For evaluation the results we use two methods. The first one is by precision (P), recall (R) and F1-measure which are widely used in Information Retrieval. For each document, the manually extracted sentences are considered as the reference summary (denoted by Summ$_{ref}$). This approach compares the candidate summary (denoted by Summ$_{cand}$) with the reference summary and computes the P, R and F1-measure values as shown in formula (8) [12].

$$P = \frac{|summ_{ref} \cap summ_{cand}|}{|summ_{cand}|} \qquad (6)$$

$$R = \frac{|summ_{ref} \cap summ_{cand}|}{|summ_{ref}|} \qquad (7)$$

$$F_1 = \frac{2PR}{P+R} \qquad (8)$$

The second measure we use the ROUGE toolkit [5], [6] for evaluation, which was adopted by DUC for automatically summarization evaluation. It has been shown that ROUGE is very effective for measuring document summarization. It measures summary quality by counting overlapping units such as the N-gram, word sequences and word pairs between the candidate summary and the reference summary. The ROUGE-N measure compares N-grams of two summaries, and counts the number of matches. The measure is defined by formula (9) [5], [6].

$$\text{ROUGE} - N = \frac{\sum_{S \in \text{summ}_{ref}} \sum_{N-\text{gram} \in S} \text{Count}_{match}(N-\text{gram})}{\sum_{S \in \text{summ}_{ref}} \sum_{N-\text{gram} \in S} \text{Count}(N-\text{gram})} \quad (9)$$

where N stands for the length of the N-gram, Count$_{match}$ (N-gram) is the maximum number of N-grams co-occurring in candidate summary and a set of reference–summaries. Count(N _ gram) is the number of N-grams in the reference summaries. We use two of the ROUGE metrics in the experimental results, ROUGE-1 (unigram-based) and ROUGE-2 (bigram-based).

### 6.3 Simulation strategy and parameters

The parameters of our method are set as follows: depth of tree that is created for any word, *n=3*; extra value for *Lesk* algorithm, $\lambda = 5$; Finally, we would like to point out that algorithm was developed from scratch in C#.net 2008 platform on a Pentium Dual CPU, 1.6 GHz PC, with 512 KB cache, and 1 GB of main memory in Windows XP environment.

### 6.4 Performance evaluation and discussion

We compared our method with four methods CRF [12], NetSum [13], Manifold–Ranking [15] and SVM [16]. Tables 1 and 2 show the results of all the methods in terms ROUGE-1, ROUGE-2, and F1-measure metrics on DUC01 and DUC02 datasets, respectively. As shown in Tables 1 and 2, on DUC01 dataset, the average values of ROUGE-1, ROUGE-2 and F1 metrics of all the methods are better than on DUC02 dataset. As seen from Tables 1 and 2 Manifold–Ranking is the worst method, In the Tables 1 and 2 highlighted (bold italic) entries represent the best performing methods in terms of average evaluation metrics. Among the methods NetSum, CRF, SVM and Manifold–Ranking the best result shows NetSum.

We use relative improvement $\frac{(\text{our method} - \text{other methods})}{\text{other methods}} \times$ 100 for comparison. Compared with the best method NetSum, on DUC01 (DUC02) dataset our method improves the performance by 2.65% (3.62%), 4.26% (10.25%) and 1.81% (3.27%) in terms ROUGE-1, ROUGE-2 and F1, respectively.

Table 1:
Average values of evaluation metrics for summarization methods (DUC01 dataset).

| Methods | Av.ROUGE-1 | Av.ROUGE-2 | Av.F1-measure |
|---|---|---|---|
| Our method | 0.47656 | 0.18451 | 0.48124 |
| NetSum | 0.46427 | 0.17697 | 0.47267 |
| CRF | 0.45512 | 0.17327 | 0.46435 |
| SVM | 0.44628 | 0.17018 | 0.45357 |
| Manifold–Ranking | 0.43359 | 0.16635 | 0.44368 |

Table 2:
Average values of evaluation metrics for summarization methods (DUC02 dataset).

| Methods | Av.ROUGE-1 | Av.ROUGE-2 | Av.F1-measure |
|---|---|---|---|
| Our method | 0.46590 | 0.12312 | 0.47790 |
| NetSum | 0.44963 | 0.11167 | 0.46278 |
| CRF | 0.44006 | 0.10924 | 0.46046 |
| SVM | 0.43235 | 0.10867 | 0.43095 |
| Manifold–Ranking | 0.42325 | 0.10677 | 0.41657 |

## 7. Conclusion

We have attacked single document summarization. our algorithm is able to select sentences that human summarizers prefer to add to their summaries. our algorithm relies on WordNet which is theoretically domain independent, and also we have used Wikipedia for some of the words that do not exist in the WordNet. For summarization, we aimed to use more cohesion clues than other lexical chain based summarization algorithms. Our results were competitive with other summarization algorithms and achieved good results. Using co-occurrence of lexical chain members, our algorithm tries to build the bond between subject terms and the object terms in the text. With implicit segmentation, we tried to take advantage of lexical chains for text segmentation. It might be possible to use our algorithm as a text segmenter.

## References

[1] Alguliev, R. M., & Alyguliev, R. M. (2007). Summarization of text-based documents with a determination of latent topical sections and information-rich sentences. *Automatic Control and Computer Sciences*, *41*, 132–140.

[2] Dunlavy, D. M., O'Leary, D. P., Conroy, J. M., & Schlesinger, J. D. (2007). QCS: A system for querying, clustering and summarizing documents. *Information Processing and Management*, *43*, 1588–1605.

[3] Erkan, G., & Radev, D. R. (2004). Lexrank: Graph-based lexical centrality as salience in text summarization. *Journal of Artificial Intelligence Research*, *22*, 457–479.

[4] Jones, K. S. (2007). Automatic summarizing: The state of the art. *Information Processing and Management*, *43*, 1449–1481.

[5] Lin, C. -Y. (2004). ROUGE: A package for automatic evaluation summaries. *In Proceedings of the workshop on text summarization branches out*, (pp. 74–81). Barcelona, Spain.

[6] Lin, C. -Y., & Hovy, E. H. (2003). Automatic evaluation of summaries using N-gram co-occurrence statistics. *In Proceedings of the 2003 conference of the north american chapter of the association for computational linguistics on human language technology (HLT-NAACL 2003)*, (pp. 71–78). Edmonton, Canada.

[7] Mihalcea, R., & Ceylan, H. (2007). Explorations in automatic book summarization. *In Proceedings of the 2007 joint conference on empirical methods in natural language processing and computational natural language learning (EMNLP-CoNLL 2007)*, (pp. 380–389). Prague, Czech Republic.

[8] Navigli, R., & Lapata, M. (2010). An Experimental Study of Graph Connectivity for Unsupervised Word Sense Disambiguation. *IEEE Computer Society , 32*.

[9] Porter, M. (1980). An algorithm for suffix stripping. *Program , 14*, 130–137.

[10] Radev, D., Hovy, E., & McKeown, K. (2002). Introduction to the special issue on summarization. *omputational Linguistics , 22*, 399–408.

[11] Salton, G., Singhal, A., Mitra, M., & Buckley, C. (1997). Automatic text structuring and summarization. *Information Processing and Management , 33*, 193–207.

[12] Shen, D., Sun, J. -T., Li, H., Yang, Q., & Chen, Z. (2007). Document summarization using onditional random fields. *In Proceedings of the 20th international joint conference on artificial intelligence (JCAI 2007)*, (pp. 2862–2867). Hyderabad, India.

[13] Svore, K. M., Vanderwende, L., & Burges, C. J. C. Enhancing single-document summarization by combining RankNet and third-party sources. *In Proceedings of the 2007 joint conference on empirical methods in natural language processing and computational natural language learning (EMNLP-CoNLL 2007)*, (pp. 448–457). Prague, Czech Republic.

[14] Wan, X. (2008). Using only cross-document relationships for both generic and topic-focused multi-document summarizations. *Information Retrieval , 11*, 25–49.

[15] Wan, X., Yang, J., & Xiao, J. (2007). Manifold-ranking based topic-focused multidocument summarization. *In Proceedings of the 20th international joint conference on artificial intelligence (IJCAI 2007)*, (pp. 2903–2908). Hyderabad, India.

[16] Yeh, J-Y., Ke, H-R., Yang, W-P., & Meng, I-H. (2005). Text summarization using a trainable summarizer and latent semantic analysis. *Information Processing and Management , 41*, 75–95.

[17] McDonald, D. M., & Chen, H. (2006). Summary in context: Searching versus browsing. ACM Transactions on Information Systems, 24, 111–141.

[18] Fung, P., & Ngai, G. (2006). One story, one flow: Hidden Markov story models for multilingual multi document summarization. ACM Transaction on Speech and Language Processing, 3, 1–16.

[19] Cilibrasi, R. L., & Vitanyi, P. M. B. (2007). The Google similarity measure. IEEE Transaction on Knowledge and Data Engineering, 19, 370–383.

**MOHSEN POURVALI** received him B.S. degree from the Department of Computer Engineering at Razi University, in 2007.Currently; he is pursuing his M.S. degree in the Department of Electrical & Computer Qazvin University. His research areas include Data mining and Text mining.

**MOHAMMAD SANIEE ABADEH** received his B.S. degree in Computer Engineering from Isfahan University of Technology, Isfahan, Iran, in 2001, the M.S. degree in Artificial Intelligence from Iran University of Science and Technology, Tehran, Iran, in 2003 and his Ph.D. degree in Artificial Intelligence at the Department of Computer Engineering in Sharif University of Technology, Tehran, Iran in February 2008. Dr. Saniee Abadeh is currently a faculty member at the Faculty of Electrical and Computer Engineering at Tarbiat Modares University. His research has focused on developing advanced meta-heuristic algorithms for data mining and knowledge discovery purposes. His interests include data mining, bio-inspired computing, computational intelligence, evolutionary algorithms, fuzzy genetic systems and memetic algorithms.